\documentclass[aps,prl,floatfix,twocolumn,preprintnumbers,amsmath,amssymb,groupedaddress,showpacs,showkeys]{revtex4-1} 

\usepackage{graphicx}  
\usepackage{dcolumn}   
\usepackage{bm}        
\usepackage{color}     
\usepackage{amsmath,amssymb}
\usepackage{psfrag}
\usepackage{bbold}
\usepackage[caption=false]{subfig}	
\usepackage{braket}  

\usepackage[usenames,dvipsnames]{xcolor}

\begin{document}
\title{What Makes Effective  Gating Possible in Two-Dimensional Heterostructures?}
\author{Predrag Lazi\'c,$^{1,2}$ Kirill D. Belashchenko,$^{3}$ and Igor \v{Z}uti\'c$^1$}
\affiliation{$^1$Department of Physics, University at Buffalo, State University of New York, Buffalo, NY 14260-1500, USA\\
	$^2$Rudjer Bo\v{s}kovi\'c Institute, PO Box 180, Bijeni\v{c}ka c. 54, 10 002 Zagreb, Croatia \\
	$^3$Department of Physics and Astronomy and Nebraska Center for Materials and Nanoscience, University of Nebraska-Lincoln, Lincoln, NE 68588-0299, USA}

\begin{abstract}
Electrostatic gating provides a way to 
obtain key functionalities in modern electronic devices and to qualitatively alter materials properties. While electrostatic description of such gating gives guidance for related doping effects,  inherent quantum properties of
gating provide opportunities for intriguing modification of materials and  unexplored devices. Using first-principles calculations
for Co/bilayer graphene, Co/BN, and Co/benzene, 
as well as a simple physical model, we show that magnetic heterostructures with two-dimensional layered materials can manifest tunable magnetic proximity effects. van der Waals bonding is identified as a requirement for large electronic structure changes by gating.
In particular, the magnitude and sign of spin polarization in physisorbed graphene can be controlled by gating, which is important for spintronic devices.
\end{abstract}

\date{\today}
\maketitle

Electrostatic modulation of carrier density is not only central to electronics, as in field-effect transistors (FETs), but
also controls
electronic phase transitions, changing an insulator into a superconductor~\cite{Goldman2014:ARMS,Ahn2006:RMP,Ye2010:NM},
or suppressing a  metal-insulator transition~\cite{Jeong2013:S}.
Such electrostatic gating is an attractive alternative to chemical doping offering reversible control of carrier density without
 altering the level of disorder,
providing new routes in obtaining desirable materials properties. A typical
range of the imposed electric field, {\bf E}$_\textrm{ext}$,
to influence electronic structures can be understood from a dielectric breakdown in air $\sim 10^{-4}$ V/{\AA} and the breaking
of a chemical bond $\sim 1$ V/{\AA}~\cite{estimate}.

While electrostatic models~\cite{Bokdam2013:PRB,Bokdam2011:NL}
show important modifications from large field-induced changes of carrier density,
the role of gating can be more complex with an inherent role of quantum tunneling effects~\cite{Brumme2014:PRB}.
Electrostatic gating can
profoundly alter magnetic materials by inducing ferromagnetism ~\cite{Zutic2004:RMP,Yamada2011:S,Ohno2000:N}
and controlling Skyrmions~\cite{White2014:PRL}.  Field-induced redistribution of carriers can
yield striking changes in the Curie temperature~\cite{Cabral2011:PRB}.

In this Letter we examine tunable magnetic proximity effects
and reveal electric control of spin-dependent properties of nonmagnetic materials. We look for  materials systems in which
gating could enable large changes in the density of states (DOS) an its spin polarization.
\begin{figure}[h]
\begin{center}
\includegraphics[clip=true,width=0.8\columnwidth]{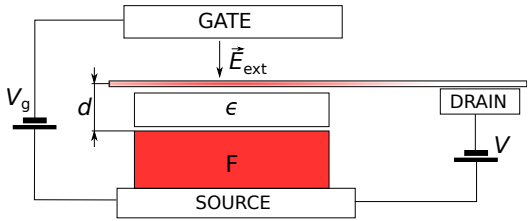}
\end{center}
\caption{Schematic of
a lateral device geometry. A ferromagnet, $F$ is separated from a graphene by a  dielectric, $\epsilon$.
The gate voltage $V_g$ and {\bf E}$_\textrm{ext}$-field are marked. The color (red) depicts
a proximity-effect induced spin polarization in graphene.
}
\label{fig:lateral}
\end{figure}

We focus on graphene (Gr)-based layered heterostructures deposited on a metallic ferromagnet (Co) as shown depicted in Fig.~\ref{fig:lateral}. The significance of this
materials choice is two-fold: (i) such systems include van der Walls (vdW) heterostructures with unique
atomically sharp
interfaces~\cite{Geim2013:N} which simplify the implementation and characterization of electrostatic gating~\cite{Ye2010:NM,Britnell2012:S},
(ii) these are key building blocks for graphene spintronics~\cite{Han2014:NN} with a prospect of gate-tunable magnetic proximity effects ---
an important precursor for lateral spin injection needed in many
applications~\cite{Zutic2004:RMP,Johnson1985:PRL,Dery2007:N,Tombros2007:N,Cho2007:APL,%
Han2010:PRL,Erve2012:NN,Dery2012:IEEETED,Lazic2014:PRB}.

\begin{figure}[h!]
\begin{center}
\includegraphics[clip=true,width=1\columnwidth]{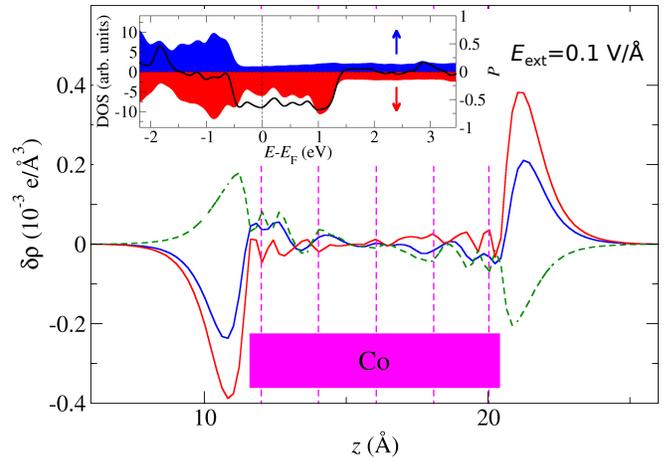}
\end{center}
\caption{(Color online) Field-induced charge rearrangement ($x-y$ plane averaged) for a 5 layer Co(0001) slab.
Majority (minority) spin: red (blue) solid lines; normalized
 magnetization: green dashed line. Vertical lines label the positions of the Co layers.
Inset: majority (minority) spin DOS: red (blue) shading; DOS spin polarization: black line.}
\label{fig:Co}
\end{figure}

Even the simple effect of electrostatic screening in a metal is intrinsically quantum mechanical, as it becomes apparent from the spin-dependent
screening in a ferromagnet~\cite{Zhang1999:PRL,Brovko2014:JPCM}, shown in Fig.~\ref{fig:Co} for a Co slab.

The electrostatic field is effectively
screened from the interior of a metal by induced surface charges of both spin projections. However, in a ferromagnet
the induced surface charge has inequivalent majority and minority 
spin contributions, leading to the changes of surface magnetic properties.  The relative amount of these charge contributions have a quantum mechanical origin: they depend
on the spin polarization of DOS, $P=(N_\uparrow-N_\downarrow)/(N_\uparrow+N_\downarrow)$,
at the Fermi level, $E_F$.  
For Co, in which $P<0$, most of the screening charge comes from the minority-spin channel~\cite{Mazin1999:PRL}.

What happens when we attach a nonmagnetic dielectric to a ferromagnet? Could its magnetic properties be altered? Unfortunately, there is a
disconnect between many
gating experiments and the lack of predictive and materials-specific methods to accurately describe them. Consequently, it is unclear what materials systems are needed to ensure that the gating will work and enable large changes in their electronic structure. Unlike conventional semiconductors in FETs, for many novel materials the corresponding screening lengths are much shorter, typically just one or two lattice constants. Electrostatic gating is therefore a surface phenomenon~\cite{Goldman2014:ARMS}
requiring an accurate description of interfaces.

\begin{figure}[t]
\begin{center}
\includegraphics[clip=true,width=1\linewidth]{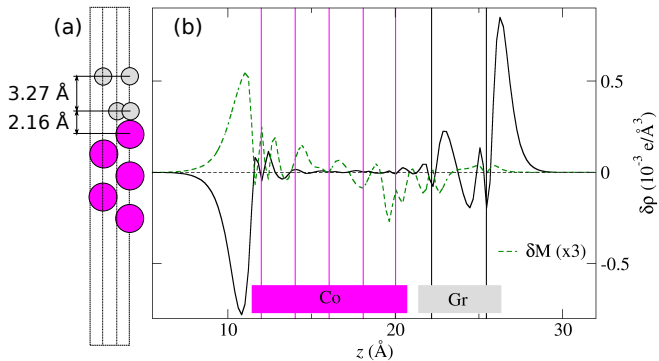}
\end{center}
\caption{(a) (Color online)  Computational geometry: 5 layer Co(0001) slab and a bilayer Gr (two C atoms/layer).
(b) Field-induced charge rearrangement ($x-y$ plane averaged):
sum for both spins (black),  3 $\times$ normalized magnetization (green).
Vertical lines: Co and Gr layers. }
\label{fig:Co2}
\end{figure}

While first-principles methods provide atomically-resolved information about heterointerfaces of interest for gating experiments, standard approaches have important limitations. 
The use of periodic boundary conditions  
introduces spurious interactions with periodically repeated images of the system which needs to be corrected~\cite{Gurel2013:JPCM,Suppl}.
To avoid these difficulties, we implement gating~\cite{Lazic2015:P} in a real-space density-functional code~\cite{Enkovaara2010:JPCM}.
For an accurate description of bonding in two-dimensional heterostructures considered here, it is crucial to include vdW interaction, which is missing in the commonly used (semi)local
functionals~\cite{Perdew1996:PRL}. This is achieved seamlessly by using vdW-DF~\cite{Dion2004:PRL}, a nonlocal correlation functional~\cite{Berland2014:X,Enkovaara2010:JPCM}.

\begin{figure*}[ht]
\begin{center}
\includegraphics[clip=true,width=1\linewidth]{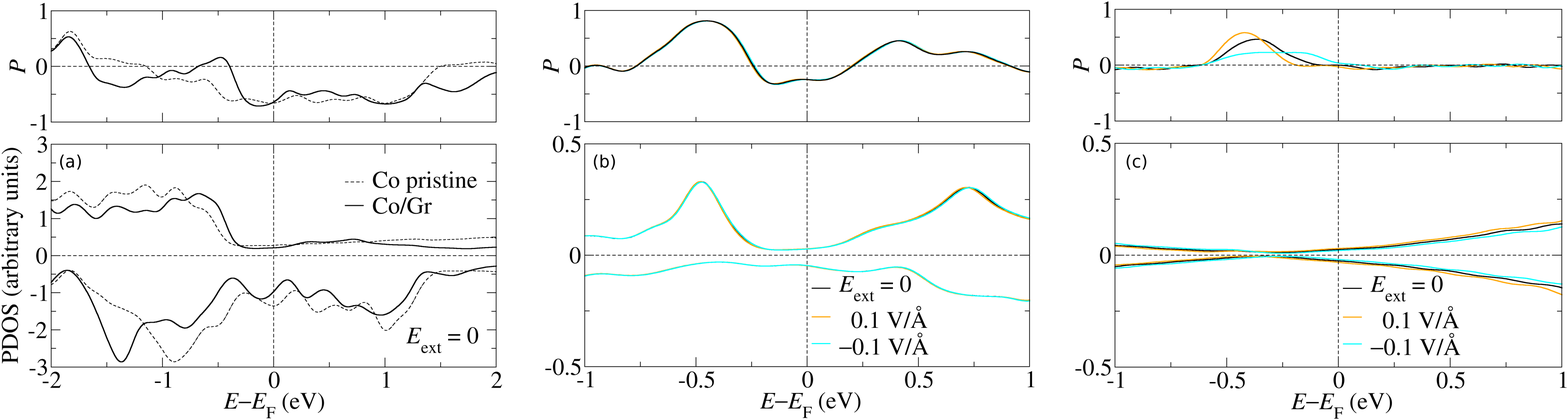}
\end{center}
\caption{
(a) (Color online) Projected DOS  (majority, minority) and the
spin polarization, $P$, on the surface Co atom  of the pristine 
slab (full) and Co slab with attached bilayer Gr (dashed).
Field-dependent PDOS and $P$ 
on the C atoms of (b) the bottom Gr layer (attached to Co surface)
and (c) top Gr layer (vdW bonded to the bottom Gr layer, chemisorbed on Co).
}
\label{fig:slab}
\end{figure*}

To explore the feasibility of gating in F/Gr heterostructures, we consider a  Co (0001) slab attached to a bilayer Gr. The bottom Gr layer
(depicted as a dielectric in Fig.~1) is chemisorbed to Co, but the top Gr layer is only weakly bound through vdW interactions. Figure 3 shows a striking difference in the response of the two Gr layers
to gating. The field-induced charge rearrangement in the top Gr layer is thrice larger compared to the bottom Gr layer. Essentially, the latter is electrically grounded through strong bonding with the Co metal. In comparison with the pure Co slab (Fig. 2), spin-dependent screening in both Gr layers is negligible. Does this mean that gating can not modify the spin polarization in a nonmagnetic region?

To answer this question, we examine the influence of the electric field on the layer-resolved DOS. As seen in Fig.~4, the adsorption of Gr changes the spin-dependent DOS of the top Co layer~\cite{note_single}.
This is consistent with the expected magnetic softening of a surface Co layer from chemisorbed Gr.
However, for both the pure Co slab and Gr-covered Co there are no field-induced changes in DOS.
In contrast to negligible DOS changes of the bottom Gr layer [Fig.~4(b)],
the changes with gating in the top layer are considerable [Fig.~4(c)].
This points to a likely trend that strongly bonded heterostructures are unsuitable for gating:
the chemical bonds ground the attached dielectric to the metallic ferromagnet (Fig.~1), precluding charge transfer and control of the spin polarization. However, the top Gr layer exhibits large field-induced changes in DOS and $P$.

Intuitively, large bonding distance could provide a large voltage drop, while small DOS suppresses screening of the external 
field $E_\textrm{ext}$.
The resulting charge transfer for the region (top Gr layer) with a small DOS at the Fermi level $N(E_F)$ will induce appreciable changes in its electronic structure. Thus, to facilitate the tunability of $P$, one should seek an energy window with a small DOS in both spin channels. In Fig.~4(c) this is observed at $E_\textrm{ext}\sim-0.4$ V/\AA \:
for the vdW-bound top Gr layer, where the Dirac cone is largely preserved.

For practical applications, it is important to ascertain that this tunability
is not unique to a specific system, and that such control of DOS and $P$ can be realized at the Fermi level. To this end, we repeated the calculations with the the bottom Gr layer replaced by hexagonal boron nitride (BN), a wide-band-gap insulator commonly used in vdW heterostructures to improve their charge and spin 
properties, as well as to implement topological states~\cite{Geim2013:N,Dean2010:NN,Wang2013:S,Guimaraes2014:PRL,Gorbachev2014:S}.

\begin{figure}[h!]
\begin{center}
\includegraphics[clip=true,width=1\columnwidth]{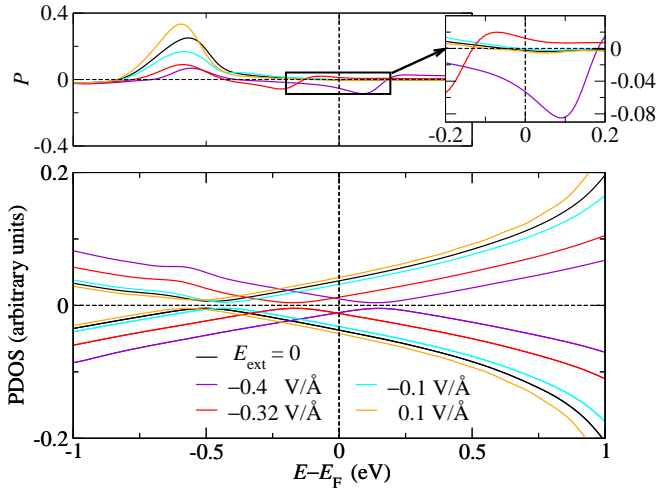}
\end{center}
\caption{(Color online)
(a) Field-dependent projected DOS and (b) the spin polarization on the C atoms of Co (slab)/ BN/Gr system.
(c) Zoom of the spin polarization in (b).
}
\label{fig:CoBNGr}
\end{figure}

The comparison of Co/Gr/Gr and Co/BN/Gr (see Supplemental Material~\cite{Suppl} for computational geometry and bonding distances)
in Figs. 5 and 4(c) shows that the replacement of the bottom Gr with BN preserves the overall DOS
shape of the top Gr layer while slightly shifting its Dirac point.
These results suggest that vdW bonding of the top Gr layer facilitates gating for two reasons:
(i) It leaves the Dirac cone largely intact, maintaining a region of small DOS; (ii) This region of small DOS can be shifted in energy by $E_\mathrm{ext}$, in contrast to chemical bonding which does not allow changes in DOS.

Although field-induced DOS changes seen in Fig.~5 resemble a (nonlinear in the field strength) rigid shift in energy, the behavior of $P$ is quite complicated.
Similar to the small DOS region in  Fig.~4(c) at $E \approx -0.4$ eV, in Fig.~5(b) we see large changes in
$P$ with $E_\mathrm{ext}$ near the Dirac point at $E \approx -0.5$ eV.
More importantly, large changes in $P$ are also observed [Fig.~5(c)] in the range of field magnitudes that bring the
Dirac point close to $E_F$. 
Surprisingly, the electric field changes both the magnitude and the sign of $P(E_F)$. This is a result of a complicated hybridization between Co, BN, and the two inequivalent C atoms (see Supplemental Material \cite{Suppl}).

While moderate $E_\mathrm{ext}$ changes [from -0.32 to -0.4 V/\AA \: in Fig.~5(c)] lead to the changes in
$P$ on the order 0.1 near $E_F$, we expect that even larger changes in spin polarization could be achieved at finite bias
in experiments on lateral transport. Similar trends of enhanced bias-dependent spin  polarization have been studied
self-consistently in magnetic {\em p-n} junctions~\cite{Zutic2002:PRL}. 

This path towards tunable magnetic proximity effect is in contrast to the common expectations that having a magnetic insulator is required to avoid a short-circuit effect of a ferromagnetic
metal~\cite{Yang2013:PRL,Wang2015:PRL,Swartz2012:ACSN}.
In our approach the first attached layer is indeed grounded to the ferromagnetic metal, but the second layer offers unexplored opportunities for spintronic devices. Even common ferromagnetic metals combined with vdW bonding could provide considerable room-temperature changes in $P$, similar to the role of MgO in achieving large tunneling magnetoresistance~\cite{Zutic2004:RMP}.

To predict the gating effects in systems similar to the one shown in Fig.~1, but not limited to Gr as the top layer, we formulate a simple electrostatic model.
We would like to estimate the shift of the DOS for Gr relative to the ferromagnet, F (the ``ground'') when an external field $E_\mathrm{ext}$ is applied by the gate. The simple calculation assumes energy-independent DOS, which is correct when the relative shift is small. We assume that charge can freely transfer between F 
and Gr over the typical time scales of the experiment (e.g. through tunneling) to establish thermodynamic equilibrium. Thus, the electrochemical potential is the same in F and Gr. When the gate voltage is applied, we then have $\delta E_F^1-e\delta V_1 = \delta E_F^2-e\delta V_2$, 
 where $\delta E_F^j$ are Fermi level shifts in the region $j$ (1: F; 
2: Gr), and $\delta V_j$ are the electrostatic potential shifts under gating. The charge density induced by the Fermi level shifts is $\sigma=eN_j \delta E_F^j$, where $N_j$ is the DOS per unit area in region $j$~\cite{note_tunnel}. 
The shifts $\delta V_j$ correspond to an induced electric  field between F and Gr: 
$\delta E_\text{in} = (\delta V_2-\delta V_1)/d$. Finally, the electrostatic boundary condition for Gr gives
$\epsilon E_\text{in}-\epsilon_0 E_\text{ext}=\sigma$, where $\epsilon_0$
is the vacuum permittivity. Combining these equations, we find, using $N_1\gg N_2$,
\begin{equation}
\delta V = \epsilon_0 E_\text{ext} d/(\epsilon+e^2 N_2 d),
\end{equation}
where $\delta V = \delta V_2-\delta V_1$ (and $\delta V_1\ll\delta V_2$). This corresponds to an effective dielectric constant
 $\epsilon_\text{eff}=\epsilon+e^2 N_2 d$.

We see an interesting interplay between classic electrostatics and quantum mechanics through the DOS of the gated material ($N_2$).
Small $N_2$ is required to achieve effective gating, while large $d$ is desirable.

\begin{figure}[t!]
\begin{center}
\includegraphics[clip=true,width=0.95\linewidth]{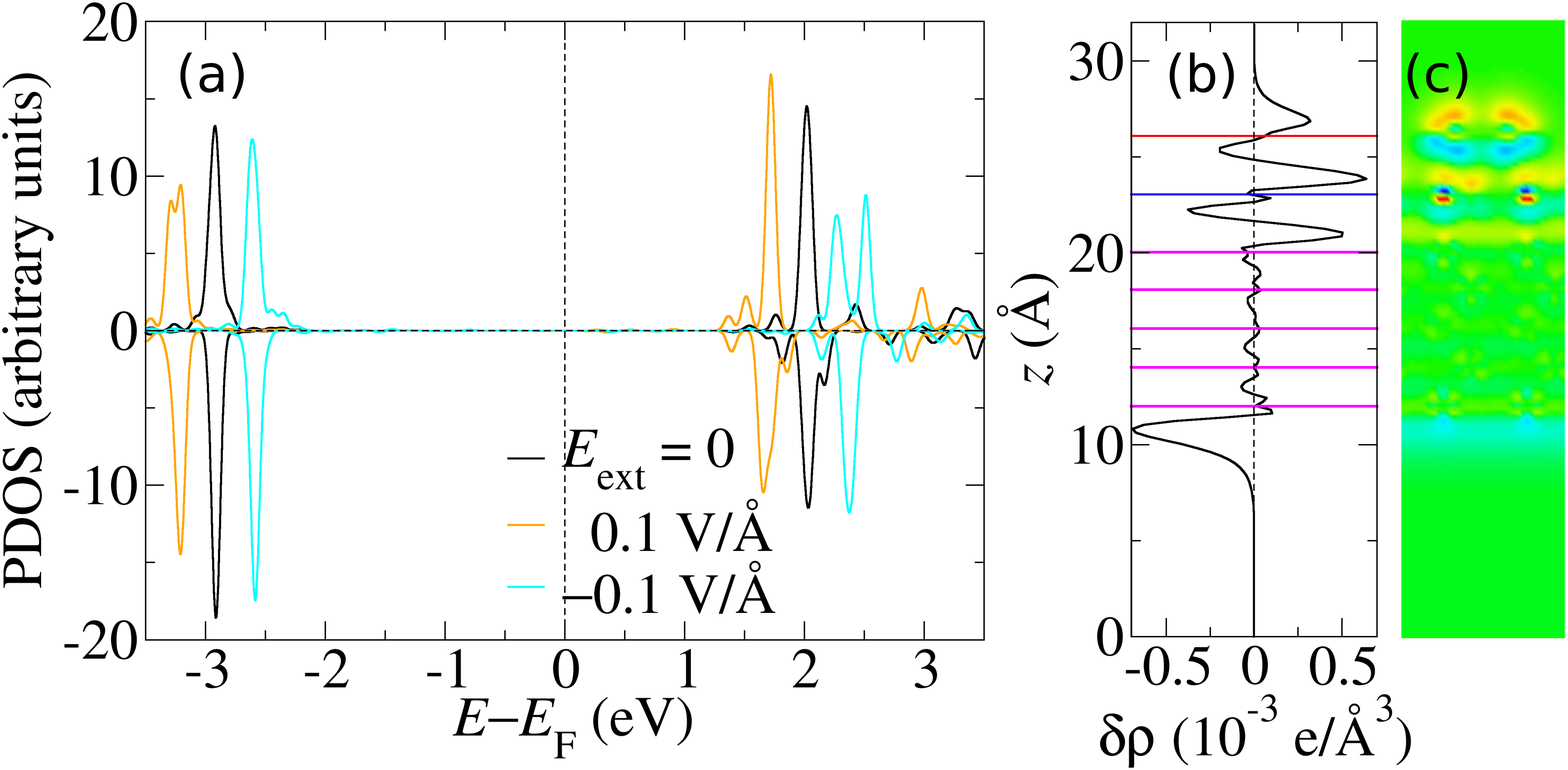}
\end{center}
\caption{(a) Field-dependent projected DOS for benzene molecule, attached to Co slab covered with BN.
(b) Surface charge transfer along the $z$ direction normal to the slab.
(c) Charge transfer at $E_\text{ext}=0.1$ V/\AA. 
The color scale represent the charge accumulation and  varies from
0.002 $e/$\AA$^3$ (red) to -0.002 $e/$\AA$^3$ (blue),
shown for a cross section through a plane.
}
\label{fig:CoBNBz}
\end{figure}

The results of the model agree well with our calculations for Co/Gr, Co/Gr/Gr, and Co/BN/Gr structures.
The DOS is an inherent property of the adsorbed top layer and a large $d$ indicates dominant vdW bonding.
These findings suggests the importance of vdW bonding for effective gating. While our model is limited to layered structures,
there are also non-layered vdW-bonded structures. For example, organic molecules are very promising building block for
spintronics~\cite{Raman2013:N}.

To examine the importance of the periodicity of the top vdW-bonded layer, we replace Gr by a non-periodic structure.
Benzene, a  dominantly vdW-bonded organic molecule, is such an example. Despite its lack of in-plane
periodicity, benzene attached to Co/BN shows in Fig.~6 a very similar behavior to all the previous cases where a periodic
monolayer (Gr) was attached. DOS changes with $E_\text{ext}$ confirm effective gating,
consistent with the prediction of the layered model [Eq.~(1)] requiring a
small DOS at $E_F$ of the adsorbed benzene, combined with a large vdW bonding distance. Moreover, the same
vdW-bonding requirement for gating applies to absorbed single atoms, as calculated for Xe~\cite{Suppl}.

A weak vdW binding (physisorption) may seem incompatible with charge transfer. However,  it is impossible to have ``pure physisorption"with no charge rearrangement. A pure vdW attraction would bring two parts of the system together
which is always prevented by the counterbalancing force due to the Pauli repulsion,  known as the ``push back'' or ``pillow''
effect~\cite{Bokdam2013:PRB,Suppl}. Such
a charge rearrangement is indeed visible in Figs.~6(b),  
(c), indicating that the vdW-bound part is inevitably slightly doped.

How could the gate-controlled $P$ be experimentally detected? Perhaps the most straightforward method is the lateral
spin transport. The changes of $P$  in Gr produced by the spin injector are detected by another, laterally-separated,
ferromagnet~\cite{Zutic2004:RMP,Johnson1985:PRL,Dery2007:N,Tombros2007:N,Cho2007:APL,Han2010:PRL,Erve2012:NN}.
It may also
be possible to directly measure proximity-induced $P$ in Gr using the magneto-optical Kerr effect (MOKE) with the single
F region in the geometry of Fig. 1. MOKE is capable of detecting very small spin polarizations, $P\sim 10^{-5}$, and
has been shown to be very sensitive in determining the structural and spin-dependent properties of Gr~\cite{Ellis2013:SR}.
A large background $P$ from F itself could be removed using double modulation scheme~\cite{Koopmans2000:PRL}
and lock-in amplifiers. Effectively, one would only detect the changes of $P$ in Gr corresponding to the chosen driven
frequency of the gate voltage (say $\sim$ 1 kHz).

While we have focused on the electric control of the magnetic proximity effect, the underlying principles of
effective electrostatic gating have broader implications. Many other systems are held together by vdW bonding, which, according to our first-principles
calculations and a simple electrostatic model, facilitates the gating of the physisorbed part of the heterostructure. This can be already
seen for gating-induced  superconductivity in vdW-bonded ZrNCl~\cite{Ye2010:NM} and we expect it to be relevant for many
vdW heterostructures, including those based on transition-metal dichalcogenides~\cite{Mak2010:PRL,Radisavljevic2011:NN}.
With dual gating structures, similar to the schematic from Fig.~1,
carrier density and the electric field can be independently changed.

\begin{acknowledgments}
P.~L. was supported by U.S. ONR Grant N000141310754. 
I.~\v{Z}. was supported by  U.S. DOE, Office of Science BES, under Award
DE-SC0004890 (theory),  NSF DMR-1124601 (applications),
and the UB Center for Computational Research.
K.\ B.\ acknowledges support from the Center for NanoFerroic Devices (CNFD), 
the Nanoelectronics Research Initiative (NRI), and NSF DMR-1308751.

\end{acknowledgments}


\begin{thebibliography}{99}
\bibitem{note_single}
Single Gr adsorbed on Co slab coincides in its electronic structure with bottom Gr in
a bilayer adsorbed on Co slab.
\bibitem{note_tunnel} 
We do not explicitly include tunneling in this model.
\end{thebibliography}
\end{document}